\pdfoutput=1
\documentclass[cits,11pt,a4paper]{article}
\usepackage{jinstpub}
\usepackage{xcolor}
\usepackage[utf8]{inputenc}
\usepackage{amsmath}
\usepackage{amsfonts}
\usepackage{amssymb}

\usepackage{graphicx}
\usepackage{epsfig}
\usepackage{tabularx}
\usepackage{lineno}
\usepackage[section]{placeins}

\title{A Compact Dication Source for Ba$^{2+}$ Tagging and Heavy Metal Ion Sensor Development}
\collaboration{The NEXT Collaboration}

\author[4,a]{K.E.~Navarro,\note[a]{Corresponding Author, karen.navarro@uta.edu}}
\author[4]{B.J.P.~Jones,}
\author[4]{J. Baeza-Rubio,}
\author[3]{M. Boyd,}
\author[3]{A.A.~Denisenko,}
\author[3]{F.W.~Foss,}
\author[4]{S.Giri,}
\author[3]{R.~Miller,}
\author[4]{D.R.~Nygren,}
\author[4]{M.R. Tiscareno,}
% ...
\affiliation[3]{
Department of Chemistry and Biochemistry, University of Texas at Arlington, Arlington, TX 76019, USA}
\affiliation[4]{
Department of Physics, University of Texas at Arlington, Arlington, TX 76019, USA}
% ...
\abstract{We present a tunable metal ion beam that delivers controllable ion currents in the picoamp range for testing of dry-phase ion sensors.  Ion beams are formed by sequential atomic evaporation and single or multiple electron impact ionization, followed by acceleration into a sensing region. Controllability of the ionic charge state is achieved through tuning of electrode potentials that influence the retention time in the ionization region.  Barium, lead, and cobalt samples have been used to test the system, with ion currents identified and quantified using a quadrupole mass analyzer.  Realization of a clean $\mathrm{Ba^{2+}}$ ion beam within a bench-top system represents an important technical advance toward the development and characterization of barium tagging systems for neutrinoless double beta decay searches in xenon gas. This system also provides a testbed for investigation of novel ion sensing methodologies for environmental assay applications, with dication beams of Pb$^{2+}$ and Cd$^{2+}$ also demonstrated for this purpose.}

\begin{document}

\maketitle

\section{Introduction}

Techniques for sensing trace levels of metal dications are of widespread importance to analytic chemistry,  biochemistry, pharmacology, environmental monitoring, and a plethora of other fields. A host of techniques are available for this purpose. Atomic absorption, inductively coupled plasma-mass spectrometry, and x-ray absorption spectroscopy are standards for low concentration detection~\cite{DingHeavy}. A major goal of emerging technologies is to prepare a sensitive and cost effective alternative to these methods, most often employing optical or electrochemical sensing strategies~\cite{LiNanostruc}. Organic turn-on chemosensors have found particular utility in biochemical microscopy where the change in fluorescence properties of organic molecules can be used to sense Ca$^{2+}$ ions at extremely low concentrations, enabling in-vivo assays of ionic activity in cells to monitor signalling dynamics in real time~\cite{SauerSingle}. Using high magnification objectives and image intensified CCD cameras, single molecule fluorescence imaging (SMFI) has been used to identify individual ions and enable super-resolution imaging, resolving features below the wavelength of the excitation light~\cite{hell20152015}.

A program initiated in 2015 within the NEXT collaboration  has focused on translating the techniques of SMFI from the solution-phase environment of biochemistry to the dry conditions of gaseous time projection chambers~\cite{nygren2016detection, jones2016single}.  If this can be achieved at large scale, it would enable the tagging of individual Ba$^{2+}$ ions emitted in neutrinoless double beta decay (0$\nu\beta\beta$) of $^{136}$Xe.  Direct observation of 0$\nu\beta\beta$  is the only known way to establish the Majorana nature of the neutrino, but searches for this process to date have been limited by background from radiogenic activity in the detector materials that effectively impose a sensitivity floor beyond which the accessible decay half-life grows extremely slowly with exposure, preventing further progress.  Identification of individual Ba$^{2+}$ ions emerging from 0$\nu\beta\beta$ could remove such radiogenic backgrounds and yield dramatic increases in sensitivity, in addition offering a confirmation mechanism for a tentative signal in a background limited experiment.  Rapid advances have been made in the development of methods to identify single barium ions or atoms in liquid and gaseous xenon for this purpose~\cite{Chambers:2018srx, mong:2014iya, rollin:2011gla, sinclair:2011zz, flatt:2007aa, mcdonald2018demonstration, rivilla2020fluorescent, jones2016single, thapa2021demonstration, thapa2020barium, Bainglass:2018odn, jones2022dynamics, herrero2022ba}.  

One persistent technical challenge is the production of a well controlled source of low energy Ba$^{2+}$ ions in vacuum or gas. The dication state is the expected charge multiplicity of barium emerging from  0$\nu\beta\beta$ in gaseous xenon, and is also the species that is amenable to fluorescence sensing with the crown ether derivatives that have been developed within NEXT for dry single molecule imaging~\cite{thapa2021demonstration, thapa2020barium, rivilla2020fluorescent}.  These molecules have been shown to enable single molecule fluorescence imaging of barium in dry conditions~\cite{thapa2021demonstration} and in high pressure gases~\cite{jones2022barium} and have been confirmed via scanning tunneling microscopy to accept ions from BaCl$_2$ salt compounds in vacuum into the computationally predicted binding sites~\cite{herrero2022ba}.  An end-to-end test of a single ion sensing scheme will require delivery of a clean and controlled flux of  Ba$^{2+}$ ions, uncontaminated by either solvent complexes or counter-ions, to a densely packed sensor layer that is monitored through SMFI.  The ion source presented in this paper was developed for this purpose.

A variety of methods exist to create ion beams, some of which have been previously used to generate Ba$^{1+}$ or Ba$^{2+}$. Plasma-driven sources such as inductively coupled plasmas (ICP) fed with analyte solutions are common ionization stages for mass spectrometers, and use of a plasma gas with ionization energy above the second ionization potential of barium is known to produce high yields of the barium dication~\cite{GodyakExpe, HoukInduc}.  Electrospray sources~\cite{Sinclair_2011} are similarly expected to produce a flux of Ba$^{2+}$ ions since this is the charge state of  barium in dissolved salts \cite{Sinclair_2011}.  Both of these sources, however, present a difficulty of removing all of the solvent molecules and counter-ions. In single ion sensing experiments, even a small quantity of solvent can spoil gas purity or interfere with capture dynamics at the otherwise dry sensing layer through solvent effects at the detection surface.  The device in this work thus employs a solution-free approach using solid source material.  

For solid analytes,  plasma discharge sputtering is commonly used for deposition of materials onto surfaces, and for fundamental studies of plasma dynamics~\cite{PandaRela, UnderwoodPhysics, KhalafCurre}.   Intense beams have been achieved by magnetically insulated pulsed diodes to create keV-energy ions ~\cite{NeriInte}.  Management of heat load  and RF noise from these sources seems to make them rather challenging for benchtop application and integration with sensitive readout systems.  Aluminosilicate ion sources have previously been demonstrated to drive ion currents of Cs$^{1+}$ in up to 640~Torr of noble gas ~\cite{AppelhansSIMION}. These thermionic ion sources mainly found in the form of natural zeolites can be ion exchanged with the cation of choice ~\cite{AppelhansSIMION, FeeneyAlum, FujisawaRecipe, OngCs+}. Our own attempts to produce a Ba$^{2+}$ aluminosilicate ion source have been unsuccessful, with the majority of ions produced in even very pure sources belonging to minority species present at the part per million level, most  notably K$^+$. Methods for barium ion production favored by barium tagging proponents in liquid xenon have included laser ablation sources, which produce Ba$^{1+}$ by focusing high energy laser pulses onto fixed solid targets~\cite{rollin:2011gla, Murray:2021sas}. Finally, radiogenic emitters have been demonstrated, using alpha decay recoils to produce Ba$^+$ ions from the alpha-unstable material coated onto barium compounds~\cite{MonteroSimple}. Neither of the latter technologies appear to generate Ba$^{2+}$ ions in significant quantities, and so they appear unsuitable for the present application. 

One method found to obtain stable production of ion beams that allows access to higher than first ionization states is vaporization followed by electron impact ionization~\cite{J-MDettmann1982, McFarlanElectron, SchroeerElectron, tinschert2012metal}.  In Ref.~\cite{popovivc2003mass}, the electron energy needed to produce Ba$^{2+}$ in crossed beam conditions was found to be around 25~eV, though this species was significantly sub-dominant to Ba$^{1+}$ in all presented conditions. Secondary electron yields of barium and other metals continue to grow rapidly with incident electron energies up to at least 300~eV~\cite{joy1995database}. Therefore, our instrument design explored electron current and energies well beyond Ba$^{2+}$ production, with an aim to breed significant concentrations of Ba$^{2+}$ ions. The device described in this work uses a modest energy thermionic electron beam (100-200 V) to generate a stable and tunable dication source, operable within a compact footprint, that will be employed in subsequent work for the testing of prototype barium tagging sensors. 

In addition to sensing of Ba$^{2+}$ ions for $0\nu\beta\beta$, the detection of small quantities of other gaseous ions in solventless conditions is also of considerable  interest.  Airborne toxic metals have been linked to mortality via moss assays~\cite{lequy2019long,wolterbeek2004atmospheric,maresca2020biological}, though a cost-effective scheme for real-time monitoring of low metal ion yields in air remains elusive.  Work is underway to develop a suitable sensing modality, mapping advances in  single molecule detection at the solid-vacuum and solid-gas interfaces that have been developed for Ba$^{2+}$ tagging in $0\nu\beta\beta$  to this problem.  Applications in ultra-high precision analytic chemistry may also be enabled by assays of metal vapors at the single molecule level.   The system described in this paper provides a versatile testbed for these studies. To this end we also demonstrate controllable ion beams formed from toxic heavy metals lead and cadmium.  Because the device couples to the sensing chamber via a single commercial vacuum flange, the system is highly flexible and transportable, enabling a wide variety of future experiments with dication beams.

\section{Apparatus \label{sec:System}}

\begin{figure}[b!]
\begin{centering}
\includegraphics[width=0.99\columnwidth]{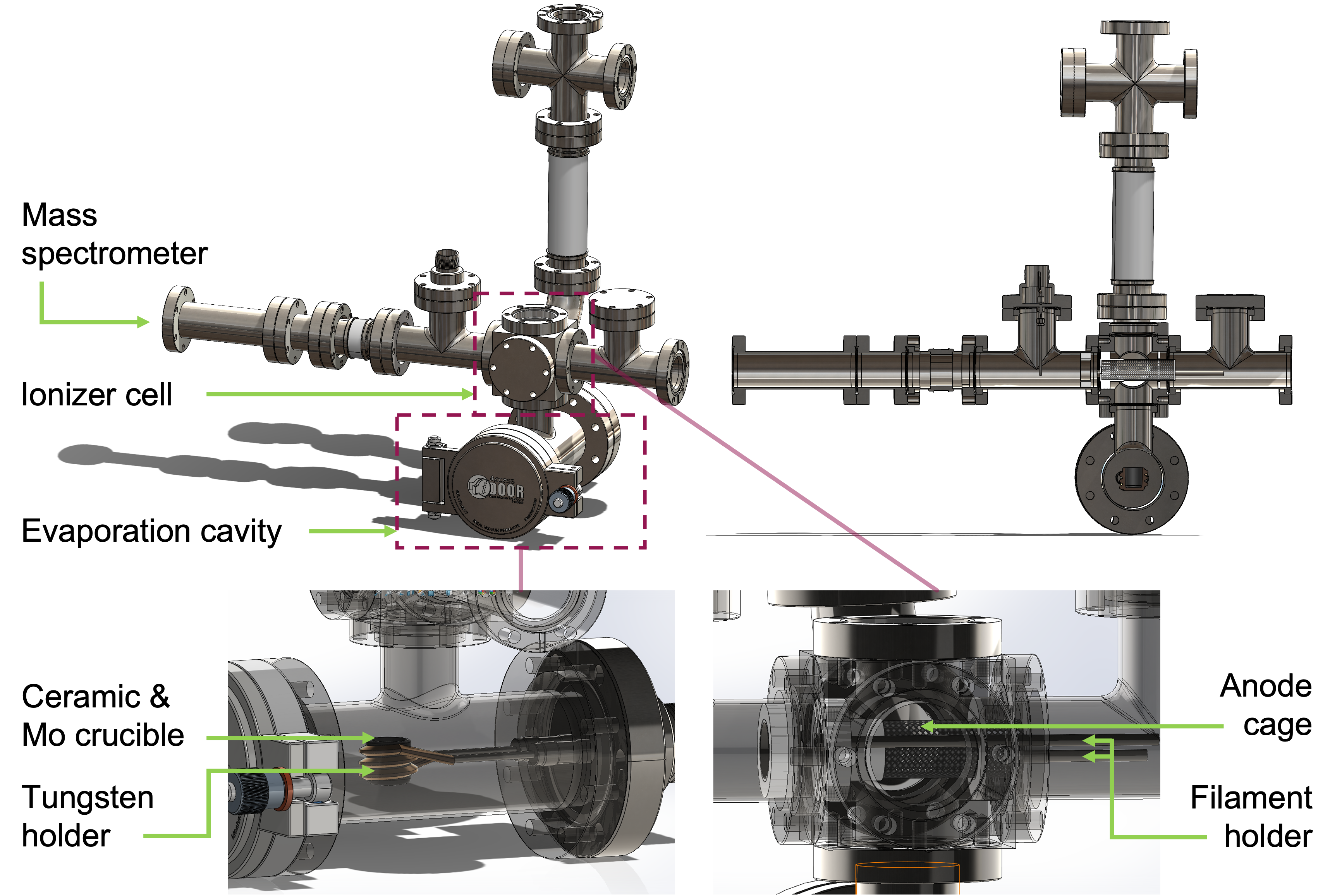}
\par\end{centering}
\caption{Diagrams of the ion beam apparatus. Top left: 3D model; Top right; front view cross section; Bottom left: evaporator detail; Bottom Right: ionizer detail.
\label{fig:Beam}}
\end{figure}

In this section we provide a detailed technical description of the dication source, which is pictured in Figure~\ref{fig:Beam}.  The device relies on metal vapor evaporation followed by multiple electron impact ionization which take place in two connected chambers. Sec.~\ref{sec:evaporator} discusses the evaporator, Sec.~\ref{sec:Ionizer} the ionizer, and Sec.~\ref{sec:MassAnalyzer} the mass analyzer.

\subsection{Evaporator\label{sec:evaporator}}

Metals for evaporation are loaded in few-gram-scale quantities into an evaporation cell at the bottom of the device, accessible through a hinged vacuum door.  For the metals used in this work, we use a molybdenum-lined ceramic crucible in order to avoid chemical attack of the ceramic by the liquefied evaporation material, which was found to erode the ceramic in the case of liquid barium. 

A central consideration driving the evaporation protocol is the maintenance of sufficiently low  vacuum to allow for electron impact ionization without absorption of the electron flux on residual gases.  Although a vacuum of 10$^{-7}$ Torr is reached before beginning experiments, heating of the crucible, copper holders and fittings raises the system temperature leading to increased out-gassing. To overcome this issue, the heat current is slowly raised while allowing the vacuum to stabilize for approximately six hours before crossing the metal liquefaction point. Handling of heavy metals is performed under a fume hood with appropriate safety precautions.  In the case of barium metal, as much of the oxide layer as possible is mechanically removed from the pellets to be evaporated before installation.  When the metal is liquefied for evaporation, any remaining oxide falls to the bottom of the liquid and does not affect the vaporization process.

When the crucible reaches the metal melting point (725~$^\circ$C for barium, 328~$^\circ$C for lead, 321~$^\circ$C for cadmium),  the metal pellets quickly liquefy.  The evaporation current of this transition varies somewhat from run to run due to an imperfectly repeatable installation geometry of the material pellets and evaporation crucible. The melting current is established visually through a glass vacuum viewport.  As soon as the material liquefies, evaporation begins.  In the case of barium, the vacuum quality is observed to rapidly improve at this point, since stray barium vapor condensing on the evaporation chamber walls serves as an excellent getter for residual water in the vacuum system.  Lead and cadmium do not exhibit this property and therefore require a better initial vacuum quality. When the crucible is operating at a stable current above the liquefaction point and the vacuum quality is better than 10$^{-6}$ Torr, the ionization electrodes and mass analyzer are switched on, to observe the resultant ion current.  A stable beam can typically be driven for at least 16 hours before the evaporation material is depleted.

The evaporation crucible is wrapped in a tungsten coil and resistively heated with up to 40~A  supplied by a high current AC supply\footnote{Kepco ATE 6-100M}.  This heating current is delivered into the vacuum chamber through a high current feedthrough with solid copper conductors and monitored via the front panel of the supply.  The metal vapor produced by evaporation is roughly beamed in the upward direction due to collimation by the walls of the 1~cm tall crucible.  A blank gasket with a 1.5~cm diameter hole between the evaporation region and the ionization region ensures that the metal vapor entering the ionization chamber is directed toward the plasma region where electron impact occurs and does not coat the inside of the ionization cell.  The evaporation chamber is cleaned of oxides and residues after each run, so that the required vacuum level can be repeatably achieved. 

\subsection{Ionizer\label{sec:Ionizer}}

Ionization takes place in a second chamber, inside a custom-built electron impact ionization source called the ionizer. The ionizer is comprised of an electron emitting filament, an anode cage that accelerates the electrons and contains the  plasma, and a series of extraction electrodes to drive  ions out of the plasma and into the detection region.  Tungsten, tantalum, iridium and nichrome filaments with different diameters were tested to find the material with the best combination of thermionic electron current, durability and stability.  Tungsten filament was found to be the superior choice, with 0.076~mm diameter wire giving the highest stable current of the various materials tested under 1~A of current supply.   The filament is 50~mm long and the current is tuned in each run to provide the required thermionic electron current flux between the filament and anode (typically 15~mA).  This latter parameter is termed the ``Electron current'' $I_e$ and is measured by an ammeter connected to the anode cage. 

The filament is negatively biased relative to the cage, which is held 5~mm radially inward from the filament by a high temperature and vacuum compatible ceramic bracket. The electrons thus accelerate inward, entering the cage with an energy corresponding to the potential difference between these electrodes. This parameter is termed the ``electron energy'', $V_e$. In vacuum conditions, each electron then follows several orbits through the ionizer before eventually being absorbed on an anode wire of the mesh cage.  When metal vapor is evaporated into the cage, the accelerated electrons can strike the metal atoms in the vapor leading to ionization, or strike metal ions in the central region leading to further impact ionization to higher charge states.

A subset of the ions are extracted from the plasma by a negatively biased extraction electrode ring at the end of the anode cage.  The potential difference between the extraction ring and anode cage influences the ion retention time in the plasma, and longer retention times generate higher charge states. A further ``lens'' electrode is placed downstream about 1~cm from the extraction electrode and can be biased to provide a focusing effect to direct the extracted beam forward.

\begin{figure}
\begin{centering}
\includegraphics[width=.99\columnwidth]{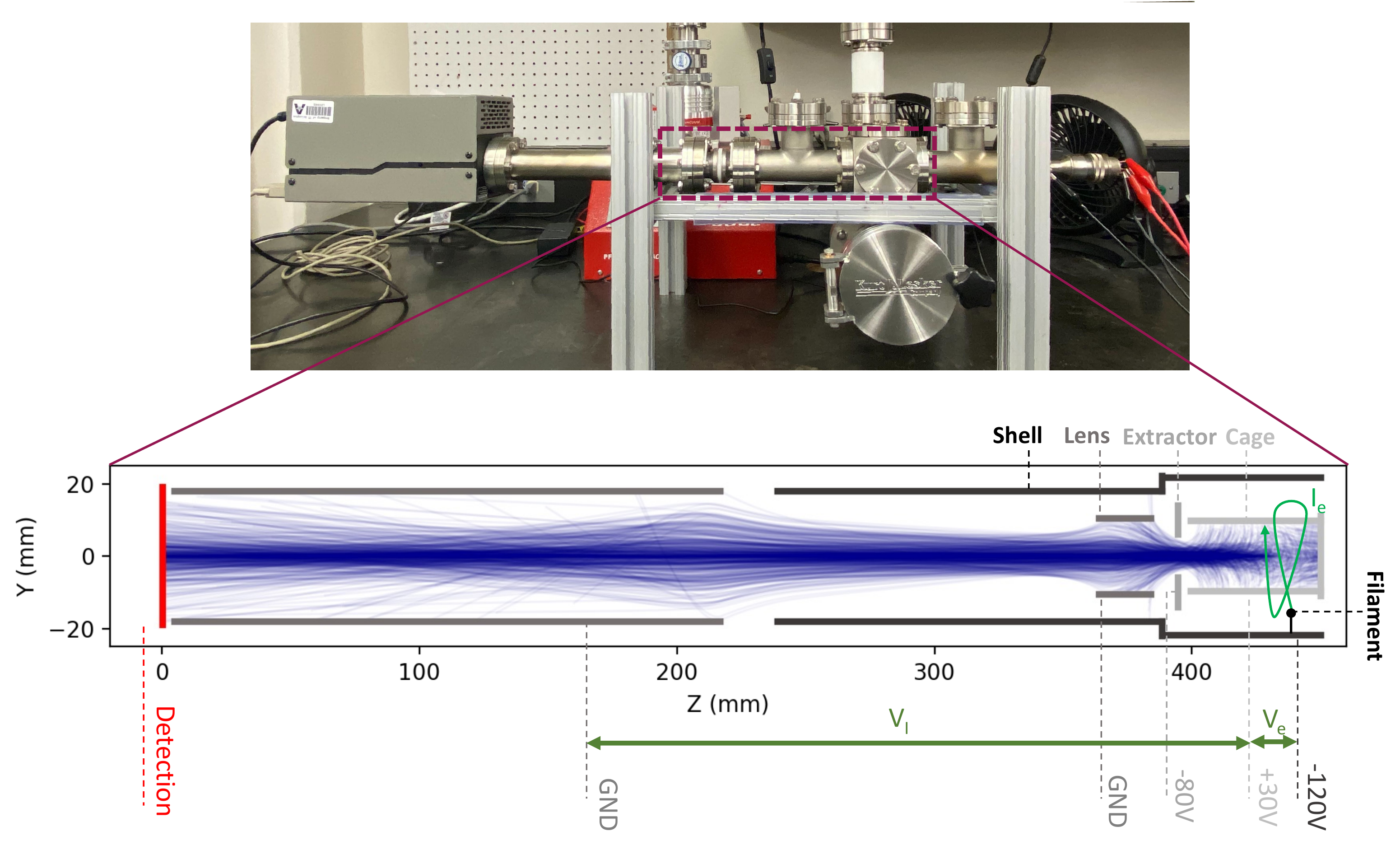}
\par\end{centering}
\caption{Photograph of the system (top) and a SIMION simulation (bottom) of 1000 ion trajectories using a representative set of voltages.
\label{fig:SimgFig}}
\end{figure}

For the studies presented in this work, the outer vacuum vessel around the ionizer region was biased at the same potential as the filament to ensure electrons travel inward to the anode cage rather than outward to the shell.   Since this shell extends part-way down the beam region before being separated from the grounded outer vacuum casing of the mass spectrometer by a dielectric break, this negative voltage also offers some focusing function.  While this design choice improves beam currents relative to those expected with a grounded outer shell around the beam region, it also means that the dependencies of beam current upon electron energy and ion transport efficiency do not factorize completely.  In practice, these effects could be separated by further fine-tuning of the various electrode potentials, independently adjusting the bias on the shell, extractor, cage, filament, focusing electrode and mass spectrometer, as well as the currents on the evaporation and electron emission filaments (a total of eight independent parameters).  The approach taken in the present study was to locate a near-optimal working point for all potentials and then study the systematic effects of three most important physical parameters $I_e$, $V_e$, and $V_I$ (the potential difference between the anode cage and ground, see Sec.~\ref{sec:MassAnalyzer}), while leaving the others fixed near the local maxima found in simulations and preliminary studies. 

\begin{figure}[t]
\begin{centering}
\includegraphics[width=.99\columnwidth]{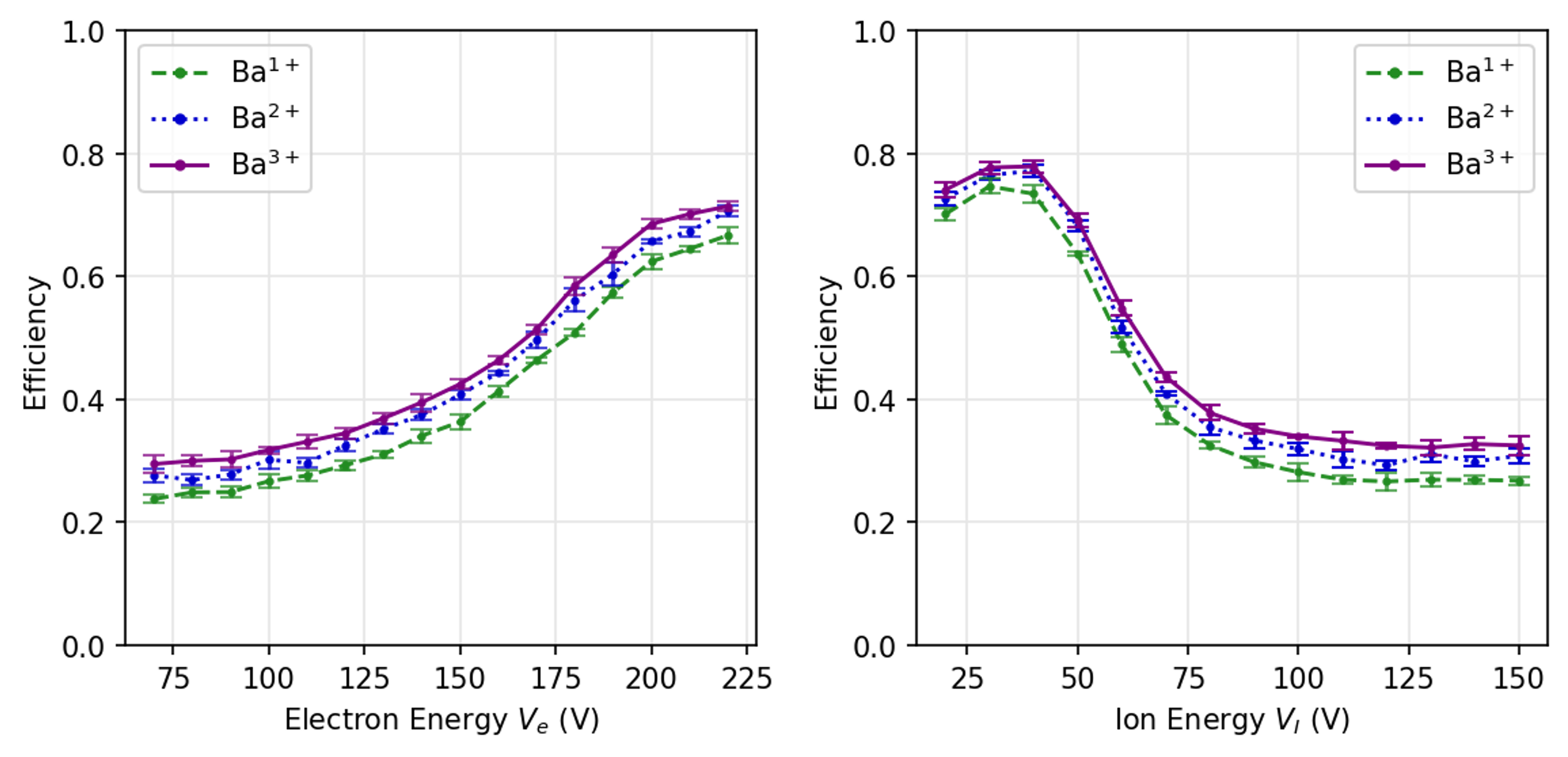}
\par\end{centering}
\caption{Simulated ion transport efficiencies for the configurations explored experimentally in this paper.
\label{fig:IonEffic}}
\end{figure}

The ion optics of the system were simulated using the SIMION software package~\cite{SIMION} and the geometries and voltages of the various electrodes were guided by these simulations.   Figure~\ref{fig:SimgFig} shows 10$^3$ simulated Ba$^{2+}$ ion transport trajectories for a representative set of electrode voltages. The ion transport efficiency, defined as the ratio of the flux of ions arriving at the detection plane to the flux of ions leaving the anode cage was evaluated over the parameter space explored in this paper and found to vary between limits of 30\% and 80\%. Figure~\ref{fig:IonEffic} shows the predicted ion transport efficiency as a function of $V_I$ at fixed $V_e=150$~eV (left) and $V_e$ at fixed $V_I = 70$~eV (right).  There is no theoretical dependence of this quantity on $I_e$, except in the limit of high electron currents where instabilities due to space charge effects become relevant.  We do not have a good technique for simulating the ion optics in this regime, but opt instead to run at electron currents below where such instabilities are observed to set in experimentally.  The predicted transport efficiencies  are sufficiently high that only modest gains are accessible by further fine tuning of ion optics, though there is some room for improvement in future work should this become necessary.

\subsection{Analyzer\label{sec:MassAnalyzer}}

The extracted ions are focused into a quadrupole mass filter\footnote{ThinkSRS RGA200 with gas ionizer disabled} with Faraday cup readout to determine the current of ions at each value of mass divided by charge (m/q).  The energy the ions have when they enter the quadrupole is determined by the potential difference between the anode cage and ground, or ion energy, $V_I$.  Notably, the true ion energy for each species is $N$ times this value, where $N^+$ is the charge state of the ion.  Some potential difference is necessary to extract the ions from the plasma and drive them toward the quadrupole, but for ion energies that are too high the quadrupole filter is expected to lose efficiency.   Experimentally, we have found that the quadrupole mass filter is still effective over the parameter space studied in this work, up to $V_I=120$~V.  At far higher ion energies the peaks begin to become smeared in the tailing direction as the quadrupole loses mass resolving power.  The studies presented in this paper are safely below this regime, so the ion currents measured at the Faraday cup are expected to be reliably measured at all parameter points.

\section{Ion Beam Analyses}

In this section we present experimental results obtained with different ion beams.  First, Sec.~\ref{sec:Xenon} discusses initial calibration with xenon that was made to test the analyzer and mass spectrometer independently of the evaporation stage at similar (m/q) to the target metals.  Sec.~\ref{sec:Ba Analysis} will discuss detailed exploration of system parameters made using a barium vapor source, a primary goal of this device. Finally  Sec.~\ref{sec:Other metals} demonstrates the application of the system to other metals, in particular lead and cadmium, which are of interest for environmental assays.

\begin{figure}
\begin{centering}
\includegraphics[width=.99\columnwidth]{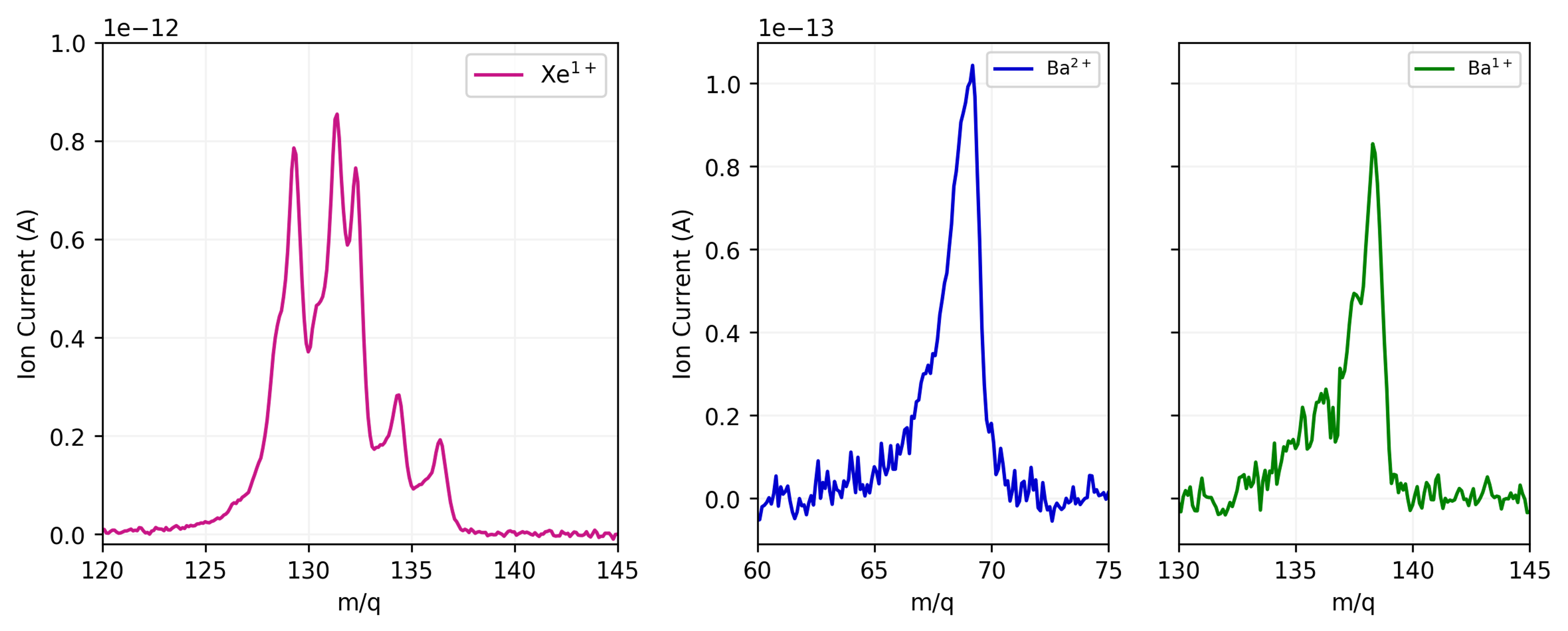}
\par\end{centering}
\caption{Left: Calibration of the system was initially performed by introducing xenon via out-gassing. The mass to-charge spectrum of Xe$^{1+}$  shows the expected natural abundance of isotopes. Right: Mass to-charge spectrum during barium evaporation showing examples of Ba$^{2+}$ and Ba$^{1+}$ peaks. 
\label{fig:Xecal}}
\end{figure}

\subsection{Ionizer calibration with xenon \label{sec:Xenon}}

Prior to installation of the metal vapor sources a first calibration was performed using xenon gas, to test the ionizer independently from the evaporator. To achieve a sufficiently low pressure of xenon in the vacuum system that electron acceleration was not compromised, xenon was introduced via outgassing from a thermoplastic reservoir.  A small PEEK block that had been previously pressurized with xenon at 4~bar in an external chamber was inserted into the evaporation region, in thermal contact with the surrounding walls. The region near the block was warmed by heat tapes to 150~$^\circ$C to cause the PEEK to outgas xenon. In this manner, neutral xenon gas was introduced into the system at a low and controllable pressure, which could be used to test the ionizer and mass analyzer.  This source was used to establish preliminary values for the various ionizer voltages and currents to be used for metal evaporation. Figure~\ref{fig:Xecal}, left shows the detected mass to-charge distribution of singly ionized Xe obtained in this calibration procedure. The individual isotope peaks are well resolved and found to be a good match to the expected natural abundance of Xe isotopes. A starting parameter point for subsequent runs with barium vapor was established at $V_e=150$~V, $V_I=70$~V,  $V_{\mathrm{extractor}}=-80$~V, $V_{\mathrm{lens}}=0$~V.

\subsection{Barium ion beam characterization \label{sec:Ba Analysis}}

\begin{figure}
\begin{centering}
\includegraphics[width=.99\columnwidth]{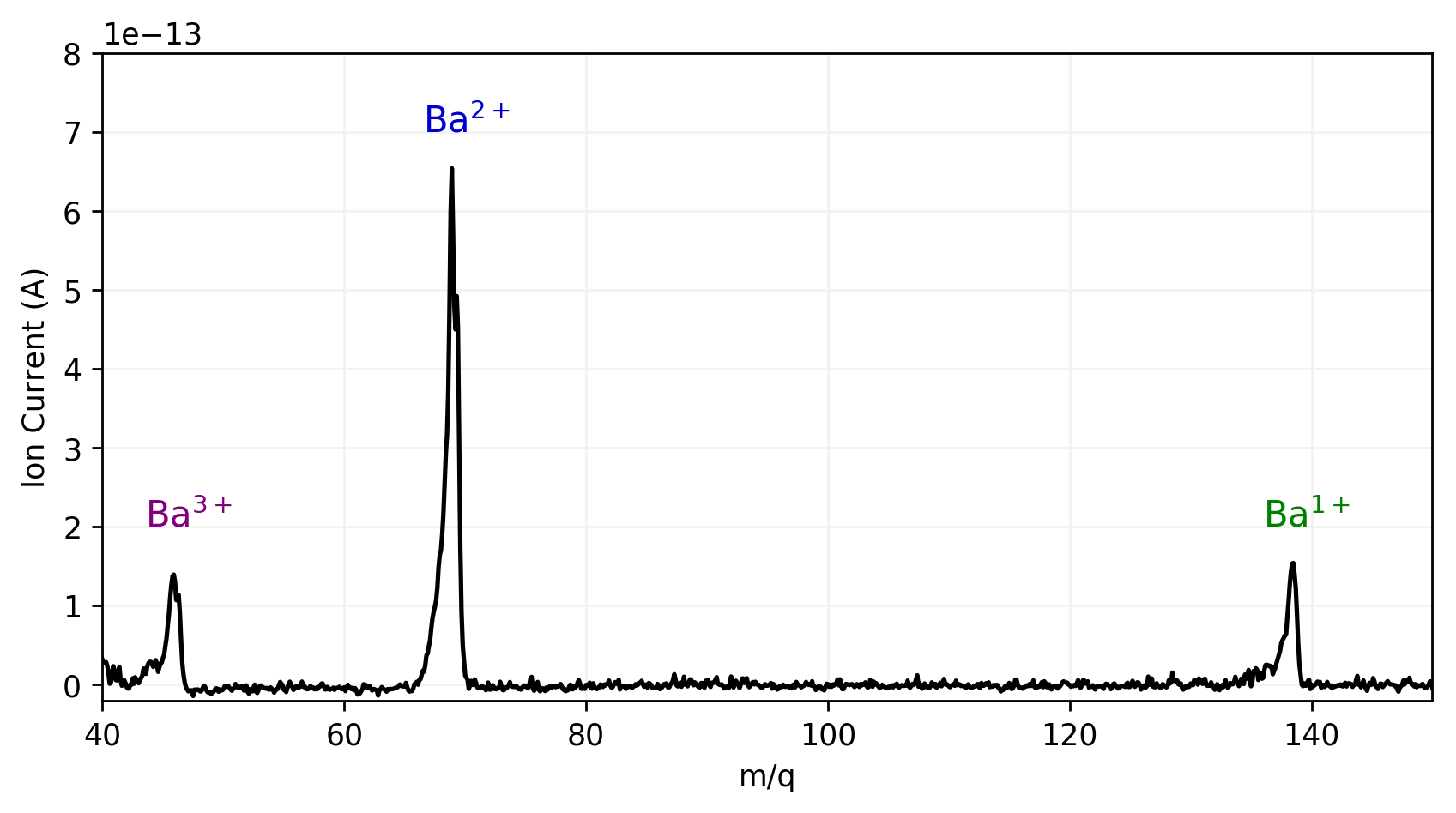}
\par\end{centering}
\caption{Complete mass to-charge spectrum from m/q values ranging from 40 to 150. Lower spectra signals correspond to residual gases in the vacuum system. The m/q values for the dominant isotope of $\mathrm{Ba^{1+}}$, $\mathrm{Ba^{2+}}$ and $\mathrm{Ba^{3+}}$ are shown.
\label{fig:Ba123+}}
\end{figure}

Barium was loaded as metal pellets into the evaporation crucible and used to produce ion beams in various system configurations. Of particular interest to us are the dependencies of the total ion yield and charge state distribution as a function of operating parameters $I_e$, $V_e$ and $V_I$. An example of a Ba$^{2+}$ spectrum is shown in Fig.~\ref{fig:Xecal}, right.  The isotopic peaks in the Ba$^{2+}$ spectrum are closer than in Xe$^{1+}$  by a factor of two, and the total fluxes in this run are lower than in the xenon calibration run, but the expected isotopic distribution is still observed. Charge states Ba$^{1+}$, Ba$^{2+}$, Ba$^{3+}$ are all clearly observed, and Ba$^{4+}$ has also been seen in some spectra with the lowest ion extraction energies.  An example wide-scan spectrum showing clearly the first three charge states is provided in Fig.~\ref{fig:Ba123+}.    One of the primary system design goals is to optimize the yield of Ba$^{2+}$ relative to other species. As will be discussed, we find that under optimal operating conditions an ion beam dominated by Ba$^{2+}$ can be obtained.

We now discuss the effects of varying the operating parameters, beginning with $I_e$.  Each dataset was taken by continuously recording the ion flux at three quadrupole settings corresponding to m/q for the leading barium isotope in each charge state, and varying the ionizer potentials in time. One minute was allowed for the system to stabilize after changing any parameter. Each data point then represents the average of data taken in 10 second intervals for each scan point over two minutes, with error bars given by the standard deviation of these measurements to illustrate the system stability.  In addition to adjusting the ionization potentials, the total ion flux can also be controlled by changing the evaporation current and hence the crucible temperature within a working range of $10^{-14}$-$10^{-11}$~A.  For the present studies we work with currents of order $10^{-13}$~A and scan over operating parameters of the ionizer.

\begin{figure}
\begin{centering}
\includegraphics[width=0.99\columnwidth]{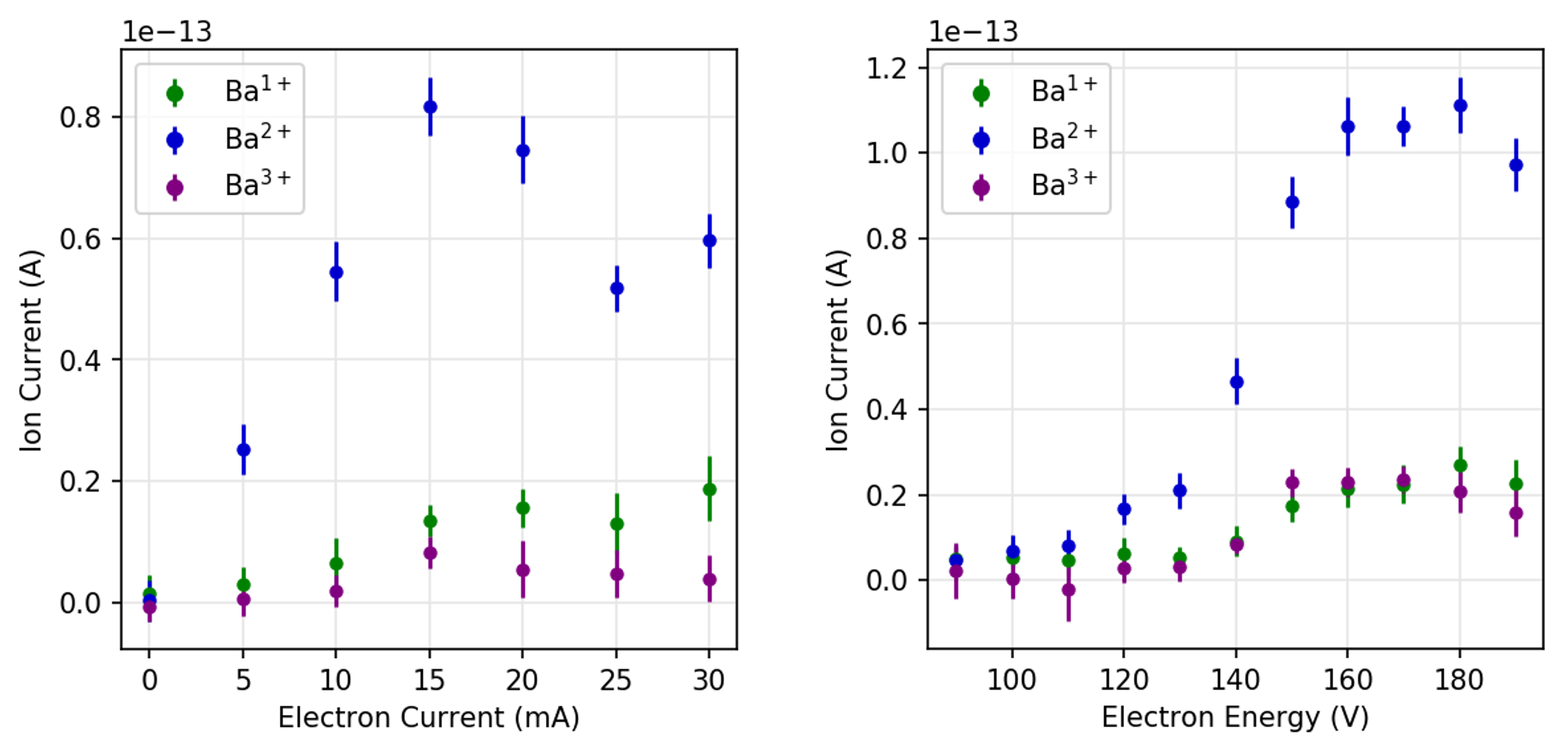}
\par\end{centering}
\caption{Studies of ion flux as a function of source parameters for barium evaporation. Left: Ion yield vs electron current $I_e$.  Right: The ion yield vs electron energy $V_e$.
\label{fig:ecurrentenergy}}
\end{figure}

Increasing  $I_e$ is expected to lead to enhanced impact ionization and a resulting increase in ion flux with an approximately linear dependence. This is indeed observed, for modest values of the electron current up to around 15~mA (Fig.~\ref{fig:ecurrentenergy}). Above this value the ionization yield appears to saturate, and also becomes somewhat unstable with time.  The threshold where this occurs depends on the values of the other source parameters, but we have found stable operation in all conditions at currents at or below 15~mA.  We attribute this behaviour to the accumulation of significant space charge from the electrons in the ionizer region, serving to inhibit extraction of the ions from the plasma. 

Increasing  $V_e$ is expected to enhance the intensity of ion beams and also, when ionization thresholds are crossed, allow for production of higher charge states.  We expect that dication fluxes should be observable for all choices of $V_e$ in our scan range, since the electron impact cross section for barium becomes non-negligible at values as low as 25~eV~\cite{popovivc2003mass}.  Due to the connection between filament bias and shell bias in our present setup, we expect a modest increase in ion fluxes in this range due to the effect of the improved ion focusing as electron energy is increased.  A larger effect that is far more difficult to model is the increase in ionization yield as the electron energy increases.  The secondary electron yield of barium metal under electron impact ionization grows quickly with incident electron energy up to at least 300~eV~\cite{joy1995database}.  These numbers have large uncertainties (a factor of 2 difference in yield separates the two  barium metal datasets given in  Ref.~\cite{joy1995database}), and are also not directly applicable to barium vapor.  Nevertheless, the mechanism that leads to this strong dependence on electron energy, that one electron of sufficient energy will ionize many metal atoms or ions with an energy-dependent cross section is active in this system.  A quantitative prediction of ion yield vs. electron energy would require detailed understanding of the electron impact cross sections on the plasma dynamics inside the ion source, and is beyond the scope of this work.  Nevertheless, a natural expectation based on both past work and SIMION simulations is a significant increase in ion yield with electron energy in the 100-200~eV range, and this is observed (Fig.~\ref{fig:ecurrentenergy}, right).  There does not appear to be a strong dependence of the charge state ratio on this parameter.

\begin{figure}
\begin{centering}
\includegraphics[width=.99\columnwidth]{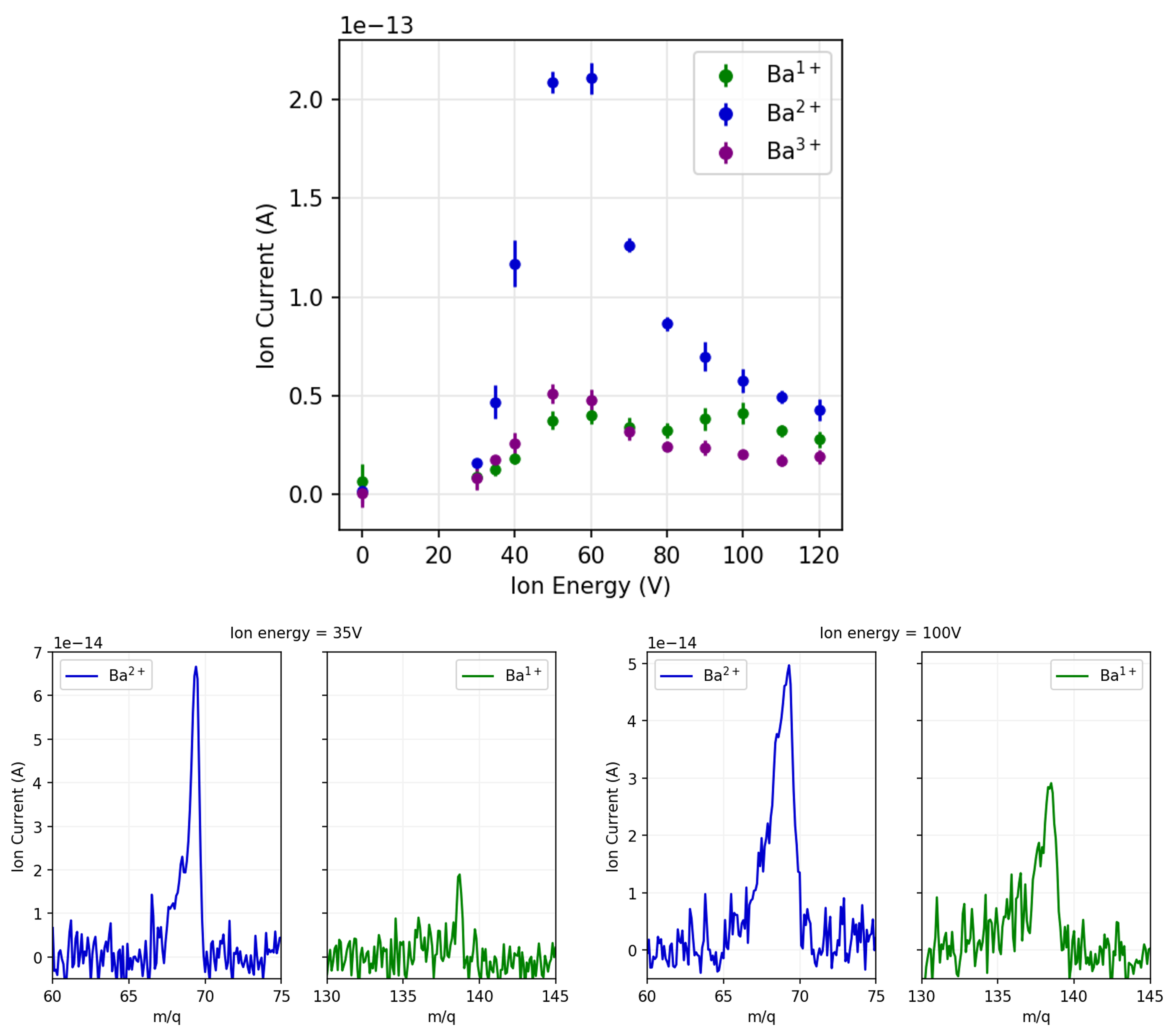}
\par\end{centering}
\caption{Top: Analysis of ion current as a function of ion energy $V_I$.  Bottom: Mass to- charge spectrum vs the ion current of barium evaporation to produce $\mathrm{Ba^{1+}}$ and $\mathrm{Ba^{2+}}$ at two different settings. Controllability of ionizer settings allow to produce a dominantly $\mathrm{Ba^{2+}}$ ion beam.
\label{fig:BigPic}}
\end{figure}

The final parameter scanned is $V_I$, the potential difference between the anode cage and ground. This is the parameter that is expected to have the largest effect on charge state distributions since it determines the ion retention time in the plasma region.  At the lowest ion energies the ions are not efficiently extracted, and at the highest ion energies we expect losses of efficiency due to the predicted falling transport probability (Fig.~\ref{fig:IonEffic}, right).  The largest ion fluxes are expected in the intermediate regime, and his is indeed observed, as shown in Fig.~\ref{fig:BigPic}.  Within the regime where the ion flux is large, lower extraction energies are expected to favor the breeding of higher charge states.  This is also observed experimentally, with the Ba$^{1+}$ to Ba$^{2+}$ ratio varying in the expected way as  $V_I$ is adjusted. In practice, because of the high electron density relative to the vapor density in our source, Ba$^{2+}$ proves to be the dominant species over all of the parameter space.

\subsection{Lead and cadmium ion beams\label{sec:Other metals}}

To demonstrate the versatility of this ion source we have also carried out investigations into production of ions from two other metals. The metals chosen where lead (Pb) and cadmium (Cd), a heavier and  a lighter element with respect to barium.  One requirement for use as an ion precursor in this system is that the material must have significant vapor pressure below around 1000~$^\circ$C.  The melting point of lead is 328~$^\circ$C  and cadmium 321~$^\circ$C, significantly lower than barium at 725~$^\circ$C. As such,  low evaporation currents were required. 

The system was prepared in the same manner as described in Sec.~\ref{sec:evaporator}, with a new molybdenum lined ceramic crucible placed inside the system and loaded with metal pellets. The system was evacuated and baked to ensure that the target pressure of p<$10^{-6}$~Torr was reached before evaporation.  The lower evaporation temperature of these two metals reduces the parasitic heating of the vacuum chamber and resulting outgassing of water from the vacuum surfaces during evaporation.  This  partially compensates for the removal of the helpful gettering action of un-ionized barium metal plating onto the walls that served to aid the vacuum quality in the barium beam runs.  

\begin{figure}
\begin{centering}
\includegraphics[width=.99\columnwidth]{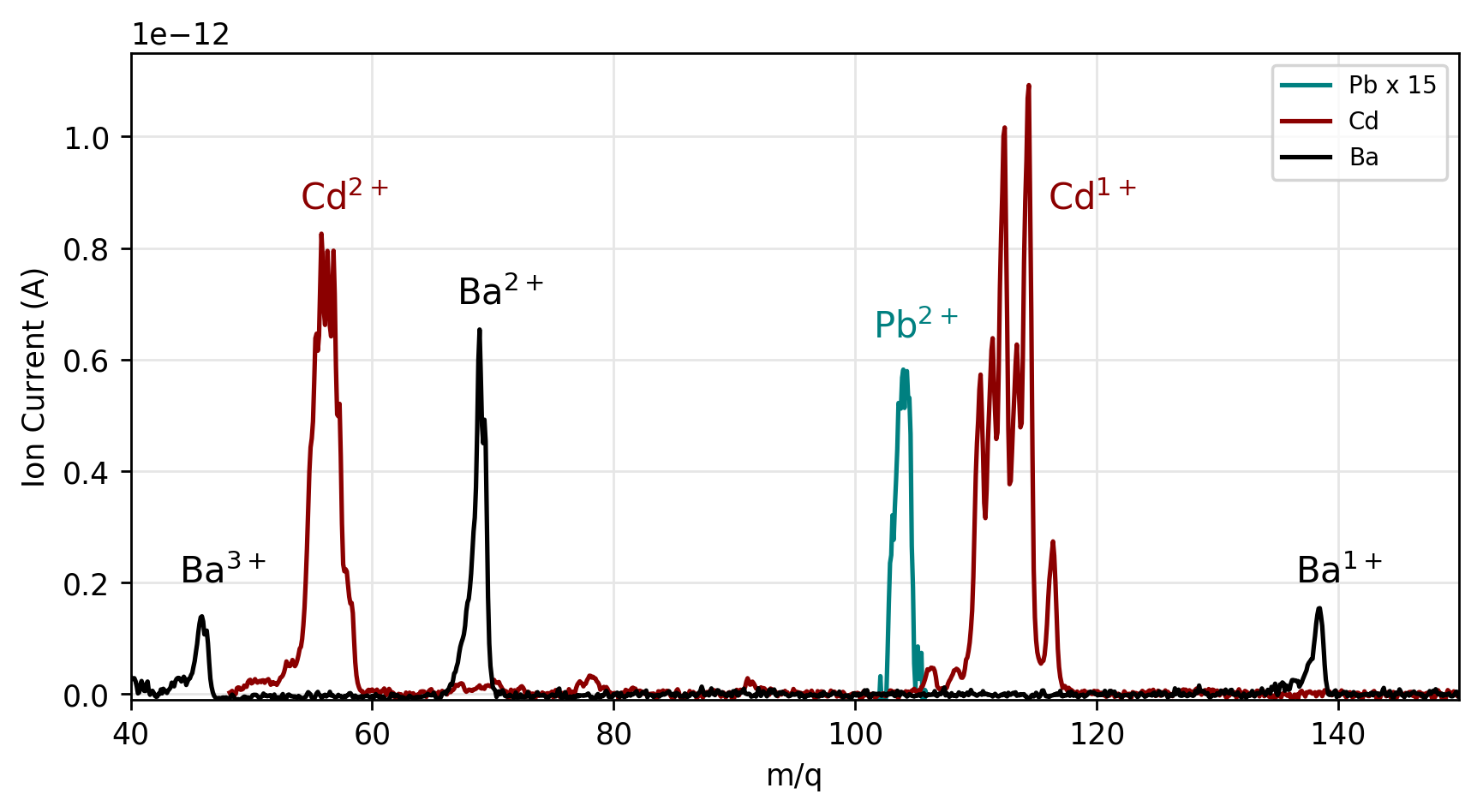}
\par\end{centering}
\caption{Mass to-charge spectrum of lead, cadmium and barium along with the charged species produced.
\label{fig:AllMetals}}
\end{figure}

Figure~\ref{fig:AllMetals} shows overlaid spectra for barium, cadmium, and lead beams, operating at the central electrode parameter point  $V_e=150$~eV, $V_I=70$~eV, $V_{\mathrm{extractor}}=-80$~V, $V_{\mathrm{lens}}=0$~V. Cadmium exhibits both singly- and doubly-charged species, similarly to barium. For lead, only the doubly charged species is observed, as the singly charged ion falls outside of the resolving range of the mass spectrometer at m/q=208.  For both Pb and Cd, the natural abundance of isotope distribution corresponds to the peaks of each spectrum.  Lower ion yields are  reported for lead, and its mass spectrum is scaled up by a factor of 15$\times$ on Fig.~\ref{fig:AllMetals} for visibility. This is because to maintain good vacuum quality while heating the filament in absence of a significant gettering action, the evaporation current was kept as low as possible while ions were still observable.  Nevertheless, useful ion currents are obtained for all species, comfortably at the level needed for testing ion sensors for the intended future applications.

\section{Conclusions\label{sec:conclusions}}

We have presented the first results from a novel ion source  designed and built to produce fluxes of metal dications in solventless conditions.  A central goal of this program is to produce a controlled flux of Ba$^{2+}$ ions for testing sensors based on single molecule imaging for barium tagging, a potentially background-free new technology to search for $0\nu\beta\beta$ in xenon gas.  Applications that build upon technologies developed for barium tagging include the sensing of toxic metal ions in atmospheric gases, and we have shown that the system is sufficiently versatile to service these applications as well, demonstrating dication beams of lead and cadmium as relevant test cases.

Using mass spectrometry we have confirmed the production of singly, doubly, and triply ionized barium with the fluxes and charge ratios exhibiting the expected dependencies on the system operating parameters including electron current, electron energy, and ion energy. By suitable selection of the system operating parameters an ion beam that is dominated by the $\mathrm{Ba^{2+}}$ charge state was achieved.   This  is the desired species for barium tagging experiments in high pressure xenon gas, as both the expected charge state emerging from $0\nu\beta\beta$ and the state that is amenable to sensing with existing SMFI-based chemosensors developed as part of the NEXT collaboration R\&D on barium tagging.  The beam is produced in entirely solventless conditions and is dominated by the Ba$^{2+}$ charge state.   Such a system, which may be augmented by additional beam cleaning stages such as conventional  E$\times$B  or quadrupole mass filtering, or by slowing or thermalization stages also under development as a part of ongoing NEXT R\&D, appears to represent an ideal source for testing dry-phase ion detection at optical chemical sensing interfaces.

\acknowledgments

This work was supported by the US Department of Energy under awards DE-SC0019054 and  DE-SC0019223, the US National Science Foundation under award number NSF CHE 2004111 and the Robert A Welch Foundation under award number Y-2031-20200401 (University of Texas Arlington).  FJS was supported by the DOE Nuclear Physics Traineeship Program award DE-SC0022359. The NEXT Collaboration also acknowledges support from the following agencies and institutions: the European Research Council (ERC) under Grant Agreement No. 951281-BOLD; the European Union's Framework Programme for Research and Innovation Horizon 2020 (2014–2020) under Grant Agreement No. 957202-HIDDEN; the MCIN/AEI of Spain and ERDF A way of making Europe under grants RTI2018-095979 and PID2021-125475NB , the Severo Ochoa Program grant CEX2018-000867-S and the Ram\'on y Cajal program grant RYC-2015-18820; the Generalitat Valenciana of Spain under grants PROMETEO/2021/087 and CIDEGENT/2019/049; the Department of Education of the Basque Government of Spain under the predoctoral training program non-doctoral research personnel; the Portuguese FCT under project UID/FIS/04559/2020 to fund the activities of LIBPhys-UC; the Israel Science Foundation (ISF) under grant 1223/21; the Pazy Foundation (Israel) under grants 310/22, 315/19 and 465; the US Department of Energy under contracts number DE-AC02-06CH11357 (Argonne National Laboratory), DE-AC02-07CH11359 (Fermi National Accelerator Laboratory), DE-FG02-13ER42020 (Texas A\&M). Finally, we are grateful to the Laboratorio Subterraneo de Canfranc for hosting and supporting the NEXT experiment.

\bibliography{Babeam.bib}

\end{document}